\def\kms{\nobreak\mbox{$\;$km\,s$^{-1}$}}
\def\la{\mathrel{\hbox{\rlap{\hbox{\lower4pt\hbox{$\sim$}}}\hbox{$<$}}}}
\def\ga{\mathrel{\hbox{\rlap{\hbox{\lower4pt\hbox{$\sim$}}}\hbox{$>$}}}}
\def\mag{\hbox{$.\!\!^{\rm m}$}}
\documentclass{kluwer}
\usepackage{graphicx}
\begin{document}
\begin{article}
% **************************************************************
\begin{opening}
\title{GAIA and the Extragalactic Distance Scale}
\author{G.A. \surname{Tammann}}
\author{B. \surname{Reindl}}
\institute{Astronomisches Institut der Universit\"at Basel \\
       Venusstrasse 7, CH-4102 Binningen, Switzerland}

\runningtitle{GAIA and the Extragalactic Distance Scale}
\runningauthor{Tammann \& Reindl}

\begin{abstract}
  The local expansion field ($v_{220}<1200\kms$) {\em and\/} the cosmic
  expansion field out to $30\,000\kms$ are characterized by
  $H_{0}=58\;$[km\,s$^{-1}$\,Mpc$^{-1}$]. While the random error of
  this determination is small ($\pm2$ units), it may still be affected
  by systematic errors as large as $\pm10\%$. The local expansion is
  outlined by Cepheids and by Cepheid-calibrated TF distances of a
  complete sample of field galaxies and by nearby groups and clusters;
  the cosmic expansion is defined by Cepheid-calibrated SNe\,Ia. The
  main source of systematic errors are therefore the shape and the
  zero point of the P-L relation of Cepheids and its possible
  dependence on metallicity. GAIA will essentially eliminate these
  systematic error sources. Another source of systematic error is due
  to the homogenization of SNe\,Ia as to decline rate $\Delta m_{15}$
  and color $(B-V)$. 
   GAIA will discover about half of the 2200 SNe\,Ia which will occur
  during a four-year lifetime within $10\,000\kms$. For many of them
  ground-based follow-up will provide useful photometric parameters
  (and spectra), which will allow to fix the dependence of the SNe\,Ia
  luminosity on $\Delta m_{15}$ and $(B-V)$ with high accuracy. At the
  same time they will yield exquisite distances to a corresponding
  number of field galaxies. --- GAIA will also revolutionize the very
  local distance scale by determining fundamental distances of the
  companion galaxies of the Milky Way and even of some spirals in- and
  possibly outside the Local Group from their rotation curves seen in
  radial velocities and proper motions.  
  
  Moreover, GAIA will obtain trigonometric parallaxes of RR Lyrae
  stars, of red giants defining the TRGB, of stars on the ZAMS, of
  White Dwarf defining their cooling sequence, and of globular
  clusters, and determine the metallicity dependence of these distance
  indicators. It will thus establish a self-controlling network of
  distance indicators within the Local Group and beyond.
\end{abstract}

\end{opening}

% **************************************************************
\section{Introduction}\label{sec:1}
% **************************************************************
%
An evaluation of GAIA's contribution to the extragalactic distance
scale requires as a first step a brief description of the present
situation (Section~\ref{sec:2}). Inadequacies of the present situation
of the distance scale are discussed in Section~\ref{sec:3}. Important
improvements expected from GAIA are outlined in
Section~\ref{sec:4}. Some conclusions are given in
Section~\ref{sec:5}.

% **************************************************************
\section{Status Quo}\label{sec:2}
% **************************************************************
%
The two pillars of the extragalactic distance scale are Cepheids and
supernovae of type Ia, which are discussed in turn, as well as other
distance indicators.
% **************************************************************
  \subsection{Cepheids}\label{sec:2_1}
% **************************************************************
%
The fundamental r{\^o}le of classical Cepheids for extragalactic
distances rests on their period-luminosity (P-L) relation, whose shape
and zero point must be known.

% **************************************************************
  \subsubsection{The shape of the P-L relation}\label{sec:2_1_1}
% **************************************************************
%
Most published Cepheid distances are based on a linear,
wavelength-dependent P-L relation as derived by Madore \& Freedman
(1991) from LMC Cepheids:
\begin{eqnarray}\label{eq:1}
   M_B & = & -2.43\,\log P - 1.13 \\ \label{eq:2}
   M_V & = & -2.76\,\log P - 1.46 \\ \label{eq:3}
   M_I & = & -3.06\,\log P - 1.87 
\end{eqnarray}
(As to the zero point see 2.1.2.).

   These equations yield apparent moduli $\mu_{B}$, $\mu_{V}$, and
$\mu_{I}$, still affected by Galactic, and in the case of
extragalactic Cepheids by internal absorption. The true distance
modulus $\mu^{0}$ can be obtained by a combination of the apparent
moduli if an absorption law is adopted (here: $A_{B}=4.1 E_{B-V}$,
$A_{V}=3.1 E_{B-V}$, and $A_{I}=1.8 E_{B-V}$), i.e.
\begin{eqnarray}\label{eq:4}
   \mu^0 & = & 4.1 \mu_{V} - 3.1 \mu_{B} \\ \label{eq:5}
   \mu^0 & = & 2.38 \mu_{I} - 1.38 \mu_{V}.
\end{eqnarray}
Almost all galaxies with Cepheid distances outside the Local Group
have been observed with HST in $V$ and $I$. If one assumes
(optimistically) that the apparent moduli $\mu_{V}$ and $\mu_{I}$ have
errors of $0\mag05$ the typical error of a Cepheid
modulus $\mu_0$ becomes $0\mag15$ (7\% in linear distance) from
equation~(\ref{eq:5}). --- If the Cepheids in a galaxy suffer variable
absorption they can be reduced one by one with equations~(\ref{eq:4})
and (\ref{eq:5}); their true distances can then be averaged.

\begin{figure}
\def\floatwidth{0.69\textwidth}
  \centerline{\includegraphics[width=\floatwidth]{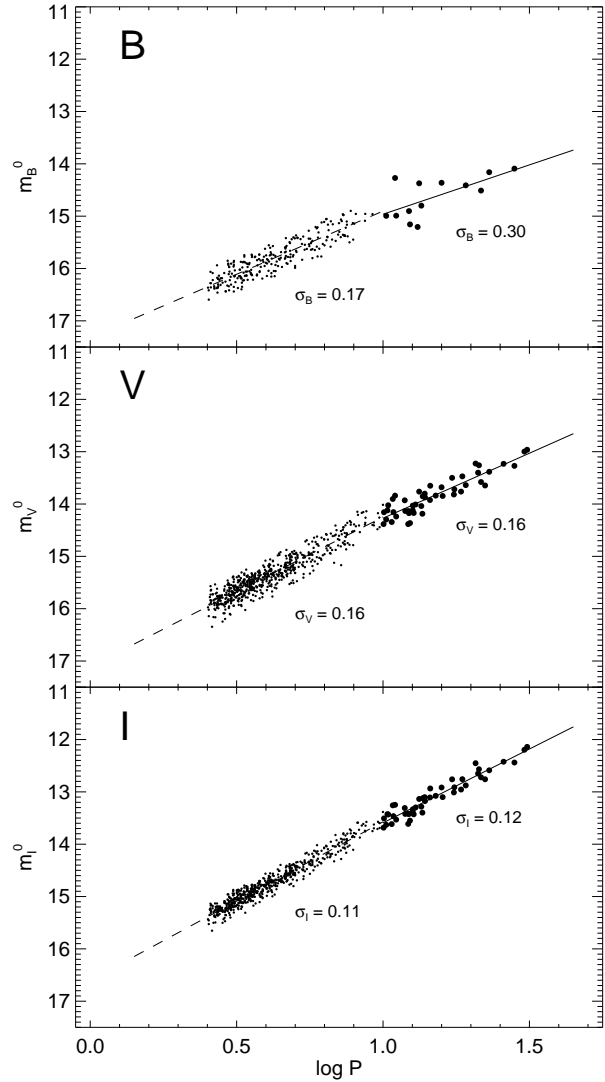}}
\caption{$P$-$L$ relations in $B$, $V$, and $I$ of the
  fundamental-mode LMC Cepheids adopted by Udalski
  et~al.\ (1999). Small symbols: $P<10^{\rm d}$; large symbols:
  $P>10^{\rm d}$.} 
\label{fig:1}
\end{figure}

\begin{figure} 
\def\floatwidth{0.7\textwidth}
  \centerline{\includegraphics[width=\floatwidth]{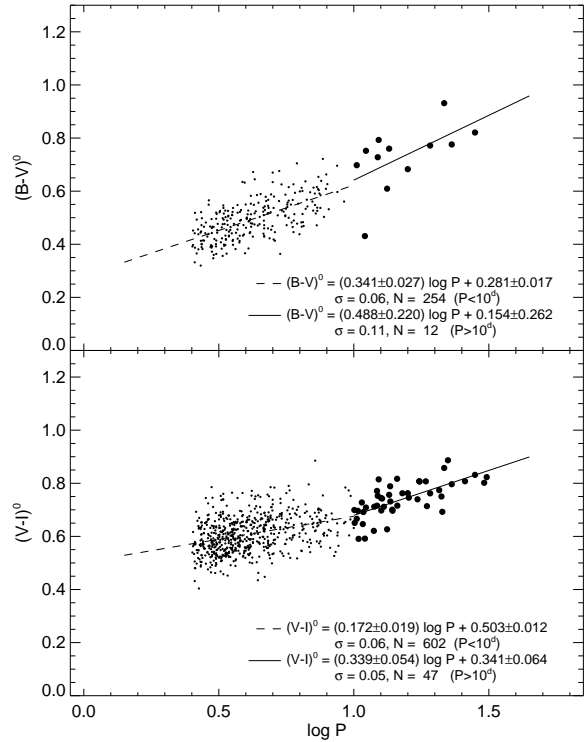}}
\caption{The $(B-V)$ and $(V-I)$ vs.\ $\log P$ diagrams of the
  fundamental-mode LMC Cepheids adopted by Udalski et~al.\
  (1999). Color excesses from that source. Symbols like in Fig.~1.}
\label{fig:2}
\end{figure}

   The OGLE programme has provided new $B$, $V$, and $I$ photometry
for hundreds of fundamental-mode LMC Cepheids (Udalski et~al.\ 1999)
which requires a significant change of slope of the P-L relations at
periods of $10\;$days %^{\rm d}$ 
(cf. Fig.~\ref{fig:1}) 
\begin{eqnarray}
\label{eq:6}
 P<10\;{\rm days}: M_{B}&=&(-2.42\pm0.08)\log P - (1.24\mp0.05); \\
\label{eq:7}
                   M_{V}&=&(-2.86\pm0.05)\log P - (1.46\mp0.03); \\
\label{eq:8}
                   M_{I}&=&(-3.03\pm0.03)\log P - (1.96\mp0.02) \\ 
\label{eq:9}
 P>10\;{\rm days}: M_{B}&=&(-1.89\pm0.62)\log P - (1.71\mp0.74); \\
\label{eq:10}
                   M_{V}&=&(-2.48\pm0.17)\log P - (1.81\mp0.20); \\
\label{eq:11}
                   M_{I}&=&(-2.82\pm0.13)\log P - (2.15\mp0.154). 
\end{eqnarray}
% ******************************************************************
%  Figure 1
% ******************************************************************
% ******************************************************************
A corresponding change of slope is also seen in the period-color (P-C)
relations (cf. Fig.~\ref{fig:2}). 
% ******************************************************************
%  Figure 2
% ******************************************************************
% ******************************************************************

Most of the available extra\,-\,Local Group Cepheid distances rely on
variables with $P>10^{\rm d}$. Their distances derived from
equation~(\ref{eq:10}) and (\ref{eq:11}) are smaller by
\begin{equation}\label{eq:12}
   \Delta\mu = -0.18 \log P + 0.14,
\end{equation}
as those from equations~(\ref{eq:2}) and (\ref{eq:3}). All following
distances are here still based on the conventional P-L relations in
equations~(\ref{eq:2}) and (\ref{eq:3}). 
% **************************************************************
  \subsubsection{The zero point of the P-L relation}\label{sec:2_1_2}
% **************************************************************
%
The zero point of the P-L relations in equations 
(\ref{eq:1})\,$-$\,(\ref{eq:3}) (as well as the one in equations 
\ref{eq:6}\,$-$\,\ref{eq:11}) is based on an adopted LMC distance
modulus of $(m-M)_{\rm LMC}=18.56$ (cf. Table~\ref{tab:1}). This value
is the straight mean of various methods to determine the LMC
distance. A straight mean is preferred here because the systematic
errors are quite uncertain. 
% ******************************************************************
%  Table 1
% ******************************************************************
\begin{table}
\caption{The Distance Modulus of LMC}
\label{tab:1}
\begin{center}
\scriptsize
\hspace*{-1cm}
\begin{tabular}{lll}
%\noalign{\medskip}
\hline%\hline
%\noalign{\smallskip}
Reference & \multicolumn{1}{c}{$\mu^0$} & Method \\
%\noalign{\smallskip}
\hline
%\noalign{\smallskip}
Sandage \& Tammann (1971) & $18.59$   & Galactic Cepheids in Clusters \\
Feast (1984) & $18.50$ & Cepheids, OB, RR\,Lyrae, MS-fitting, Mira \\
Feast \& Catchpole (1997) & $18.70\pm0.10$ & HIPPARCOS parallaxes of
Galactic Cepheids \\
Van Leeuwen et~al.\ (1997) & $18.54\pm0.10$ & HIPPARCOS parallaxes of
Miras \\
Madore \& Freedman (1998) & $18.44-18.57$ & HIPPARCOS parallaxes of
Galactic Cepheids \\
Panagia (1999) & $18.58\pm0.05$ & Ring around SN\,1987A \\
Pont (1999) & $18.58\pm0.05$ & HIPPARCOS parallaxes of Galactic Cepheids \\
Walker (1999) & $18.55\pm0.10$ & Review \\
Feast (1999) & $18.60\pm0.10$ & Review \\
Walker (1992), Udalski et~al.\ (1999) & $18.53\pm0.08$ & $M_{V}^{\rm
  RR}=0.41\pm0.07$ Sandage (1993), \\ 
       & & Chaboyer et~al.\ (1998) \\
Groenewegen \& Oudmaijer (2000) & $18.60\pm0.11$ & HIPPARCOS
parallaxes of Galactic Cepheids \\
Kov{\'a}cs (2000) & $18.52$ & Double mode RR Lyr \\
Sakai et~al.\ (2000) & $18.59\pm0.09$ & Tip of RGB \\
Cioni et~al.\ (2000) & $18.55\pm0.04$ & Tip of RGB \\
Romaniello et~al. (2000) & $18.59\pm0.09$ & Tip of RGB \\
Groenewegen \& Salaris (2001) & $18.42\pm0.07$ & Eclipsing binary \\
Girardi \& Salaris (2001) & $18.55\pm0.05$ & Red clump stars \\
Baraffe \& Alibert (2001) & $18.60-18.70$ & Pulsation theory of
Cepheids \\
%\noalign{\smallskip}
\hline
%\noalign{\smallskip}
Adopted & $18.56\pm0.02$ & \\
%\noalign{\smallskip}
\hline
\end{tabular}
\end{center}
\end{table}
% ******************************************************************

% **************************************************************
  \subsubsection{Cepheids and the local expansion field}\label{sec:2_1_3}
% **************************************************************
%
Within the Local Group Cepheid distances have been determined for
M\,31, M\,33, NGC\,6822, IC\,1613 and a few additional irregular
galaxies. Beyond the Local Group one has Cepheid distances for 31
galaxies as compiled in Tammann et~al.\ (2001), the bulk of those
being observed with HST. The subset of 23 galaxies outside the noisy
Virgo region, i.e. $\alpha_{\rm M87}>30^{\circ}$, define a Hubble
diagram out to $v_{220}\approx1500\kms$\footnote[2]{The $v_{220}$
  velocities are corrected for a self-consistent Virgocentric infall
  model with a local infall vector of $220\kms$ (Kraan-Korteweg 1986).} 
with a scatter of $\sigma_{\mu}=0\mag43$ (Fig.~\ref{fig:3} below),
part of which is due to errors of the Cepheid distances, the other
part must be attributed to peculiar motions.

% **************************************************************
  \subsection{Other Distance Indicators}\label{sec:2_2}
% **************************************************************
%
% **************************************************************
  \subsubsection{RR Lyrae stars}\label{sec:2_2_1}
% **************************************************************
%
The RR Lyrae stars are important to check the distance of LMC and SMC
and other late- {\em and\/} early-type companions of the Milky
Way. The calibration of their luminosity is difficult because it
depends on metallicity, i.e.
\begin{equation}\label{eq:13}
   M_{V} = a + b \,\mbox{[Fe/H]}.
\end{equation}
A thorough discussion of the calibration through main-sequence fitting
of globular clusters, the Cepheid distance of LMC, and the
Baade-Wesselink-Becker method has led Chaboyer et~al.\ (1998) to adopt
$a=0.39\,(\pm0.1)$ at $\mbox{[Fe/H]}=-1.9$ and  $b=0.23\pm0.04$.

% **************************************************************
  \subsubsection{The Tip of the Red Giant Branch (TRGB)}\label{sec:2_2_2}
% **************************************************************
%
The TRGB in the $I$-band has proved to be a useful (nearly)
metal-independent distance indicator out to $\sim\!10\;$Mpc and
eventually to $20\;$Mpc. For a compilation of 11 available TRGB
distances see Kennicutt et~al.\ (1998). The position of the TRGB at
$M_{I}=-4.0\pm0.1$ depends on the adopted distances of globular
cluster (Da Costa \& Armandroff 1990). The method is promising because
it can be applied to early- {\em and \/} late-type
galaxies and the observational effort is smaller than for Cepheids, but
giving about equal distance errors of $\sim\!0\mag1$-$0\mag2$.

   Also Red Giant Clump stars have been proposed as distance
indicators, but their position is metal- and age-dependent, and
their application is complex (Girardi \& Salaris 2001).

% **************************************************************
  \subsubsection{Tully-Fisher distances of field galaxies}\label{sec:2_2_3}
% **************************************************************
%
The Tully-Fisher (TF) relation, i.e.\ the correlation of the 21\,cm
(or any optical) line width $w$ (taken at 20\% maximum intensity as a
measure of the maximum rotation velocity) of a spiral galaxy with its
luminosity, has found a wide application. Its latest form is given by
(Tammann et~al.\ 2001)
\begin{equation}\label{eq:14}
  M_{\rm B}^0=-7.31 \log w_{20} - (1.833\pm0.095); \quad \sigma_{\rm M}=0.53,
\end{equation}
where the slope is taken from a {\em complete\/} volume-limited sample
of (favorably inclined) Virgo cluster spirals, while the zero point
depends entirely on 29 Cepheid distances; most of the galaxies were
observed during an HST Key Project (cf.\ Freedman et~al.\ 2001).

   Equation~(\ref{eq:14}) can be applied to a {\em complete\/} sample
of 154 spirals with $v_{220}\le1000\kms$ and inclination
$i>45^{\circ}$ (Federspiel 1999). The resulting TF distance moduli of
the 92 galaxies outside the Virgo region ($\alpha_{\rm
  M87}>30^{\circ}$) are shown in the distance-calibrated Hubble
diagram of Fig.~\ref{fig:3}. Since the dense Virgo region
($\alpha_{\rm M87}<30^{\circ}$) proper has large peculiar motions, the
dynamically defined distance limit becomes fuzzy here causing large
scatter and systematic selection effects. 

   Even outside the Virgo region the scatter about the Hubble line is
as large as $\sigma_{(m-M)}=0\mag86$.
Peculiar motions introduce a scatter of $\sigma_{(m-M)}<0\mag43$ as
judged from Cepheids (Section~2.1.3), and additional scatter is
introduced by observational errors of the line widths $w$, which were
compiled from various sources, and the
inclination-dependent absorption corrections. But it is clear, that
the {\em intrinsic\/} scatter of the TF distances cannot be smaller
than $\sigma_{(m-M)}\sim0\mag5$ (cf. equation~\ref{eq:14}). This large
scatter introduces severe selection bias (Malmquist effect), --- a
problem well known to workers with trigonometric parallaxes. If the
above sample of 92 galaxies is cut by an additional apparent-magnitude
limit, the scatter decreases and the value of $H_0$ increases; the
brighter the magnitude cut the more severe the effect becomes
(cf. Tammann et~al.\ 2001, Fig.~6). This explains why claims in the
literature for an unrealistically small scatter of the TF relation ---
as derived from incomplete samples --- are always accompanied by too
large values of $H_0$.

   The $1000\kms$ sample of TF distances is so far the deepest
complete, distance-limited sample of field galaxies for which TF data
are available. Sophisticated methods have therefore been developed to
correct {\em magnitude-limited\/} samples for Malmquist bias (Theureau
et~al.\ 1998; Federspiel et~al.\ 1998; Hendry 2001). But the results
become less convincing as the samples thin out with increasing
distance.

   It has also been proposed to determine distances of field galaxies
through the D$_{\rm n}-\sigma$ (or fundamental-plane) method, surface
brightness fluctuations (SBF) and brightest planetary nebulae. The
latter method is confined to early-type galaxies because of the
confusing effect of HII regions in late-type galaxies. The same
restriction holds for the two first-mentioned methods, unless one
assumes that the bulges of spiral galaxies follow the same relations
as early-type galaxies.
This raises the problem of calibration: there is no normal E/S0 galaxy
within 10$\;$Mpc and correspondingly none with a primary distance
determination. A meaningful calibration must therefore be based on an
{\em adopted\/} Virgo cluster distance and on a fair and sufficiently
large sample of cluster members to beat the considerable depth effect of
the cluster. An equally severe problem is that no complete
magnitude-limited, let alone distance-limited sample of D$_{\rm
  n}-\sigma$, SBF, or planetary nebulae distances is available at
present. The true intrinsic scatter of these three methods is
therefore unknown, and the distances so far available cannot be
assigned a confidence interval.

   An interesting second distance ladder may emerge, based solely on
Population~II objects. This depends on the supposed universality of
the peak of the bell-shaped luminosity function of globular clusters
(GCLF). The method has its successes and failures (Tammann \& Sandage
1999 and references therein). The failures may be due to the
difficulty to observe the faint descending branch of the GCLF, but on
the other hand the method has no physical basis and is purely
heuristic. The calibration of the peak luminosity, i.e.\
$M_{B}=-6.93\pm0.08$, rests at present on the RR~Lyrae distances of
Galactic GC and on the Cepheid distance of M\,31. The latter can be
substituted by the TRGB distance, which is almost identical (Kennicutt
et~al.\ 1998), but calibrated by old objects. 

% **************************************************************
  \subsection{Supernovae of Type Ia}\label{sec:2_3}
% **************************************************************
%
A complete sample of 26 SNe\,Ia can be established with good CCD
photometry in $B$, $V$ (and $I$) near maximum. They fulfill the
additional conditions of having $1200<v_{220}\la30\,000\kms$, to be blue
at maximum, i.e.\ $(B-V)\le0\mag06$, to have Branch-normal spectra as
far as known (Branch, Fisher \& Nugent 1993),
and to be corrected for Galactic absorption by less than $0\mag2$ (to
minimize errors in the absorption correction). If the maximum
magnitudes in $B$, $V$, and $I$ of these SNe\,Ia are homogenized as
for decline rate $\Delta m_{15}$ and color $(B-V)$ (for details see
Parodi et~al.\ 2000), they define Hubble diagrams of the form
\begin{equation}\label{eq:15}
   \log v = 0.2\,m^{\rm corr}_{B,V,I} + c_{B,V,I}
\end{equation}
with
\begin{equation}\label{eq:16}
   c_{B}=0.662\pm0.005, \quad c_{V}=0.661\pm0.005, \quad  c_{I}=0.604\pm0.005.
\end{equation}
The scatter amounts in all three relations to only $\sigma_{\rm
  m}=0\mag11$!
This small ``Hubble'' scatter is unparalleled. It can be fully
accounted for by observational errors and by a (smaller)
contribution from peculiar motions. The intrinsic luminosity scatter
of SNe\,Ia, once they are corrected for decline rate and color, is
therefore below the present detection limit. SNe\,Ia are therefore
the best standard candles known. A single SNe\,Ia yields the distance
of any galaxy to within $\pm0\mag11\;$(5\%), i.e. even better than a
sample of Cepheids. 
The small luminosity scatter of SNe\,Ia makes them at the same time
practically immune to selection bias.

   Consequently the problem of the large-scale distance scale can be
solved by equation~(\ref{eq:15}) if only the absolute magnitude of
SNe\,Ia is known. For this purpose a specific HST programme has been
mounted to determine Cepheid distances of the nearby galaxies which
have produced a SNe\,Ia. To date nine Cepheid-calibrated SNe\,Ia are
known (Saha et~al.\ 2001). Their mean absolute magnitudes are
\begin{equation}\label{eq:17}
   <\!\!M_{B}^{\rm corr}\!\!>=-19.56\pm.07, \;
   <\!\!M_{V}^{\rm corr}\!\!>=-19.53\pm.06, \;
   <\!\!M_{I}^{\rm corr}\!\!>=-19.25\pm.09.
\end{equation}
The large-scale value of $H_0$ (out to distances of
$30\,000\kms$) follows by transforming equation~(\ref{eq:15}) to
\begin{equation}\label{eq:18}
   \log H_0 = 0.2\,M_{B,V,I} + c_{B,V,I} + 5
\end{equation}
and by inserting $M$ and $c_{\lambda}$ from equations~(\ref{eq:17})
and (\ref{eq:16}), respectively, to yield a mean value of
$H_{0}(B,V,I)=56.6\pm2.3$. The linear fit to the SNe\,Ia data
corresponds closely to a $\Omega_{\rm Matter}=1$ model. If a
$\Omega_{\rm Matter}=0.3$, $\Omega_{\Lambda}=0.7$ model is preferred
$H_0$ is slightly increased to $57.4\pm2.3$ (cf. Carroll et~al.\ 1992).
The individual distances of the sample of $26+9$\footnote[2]{The 9
  SNe\,Ia have a Galactic absorption correction of $A_{V}>0\mag2$ and
  are not used for the calibration in equation~(\ref{eq:16}).} SNe\,Ia are
determined from their observed maximum magnitudes $m_{B,V,I}$ and from
equation~(\ref{eq:17}). They are plotted in Fig.~\ref{fig:3} (below).

% **************************************************************
  \subsection{Distances of Local Groups and Clusters}\label{sec:2_4}
% **************************************************************
%
It can be expected that local galaxy groups and clusters delineate the
Hubble flow particularly well because (1) their distances can be based
on several distance determinations, and (2) the {\em mean\/} velocity
of the cluster members minimizes the peculiar motions. A
self-explanatory list of seven local (and not so local) group/cluster
distances is given in Table~\ref{tab:2}. Leo has two, Fornax three
excellent SNe\,Ia distances; the Virgo distance from three SNe\,Ia is
not quite as secure because of the considerable depth of the
cluster. Six entries in Table~\ref{tab:2} include distances based on
one or more galaxies with Cepheids. The TF distance of Virgo is based
on a complete sample of 49 suitable spirals; the TF sample of UMa
may not be complete and the resulting distance is rather a lower
limit.

% ******************************************************************
%  Table 2
% ******************************************************************
\begin{table}
\caption{Distances of local groups and clusters}
\label{tab:2}
\begin{center}
\begin{minipage}{0.95\textwidth}
\scriptsize
\hspace*{-0.5cm}
\begin{tabular}{lrrl}
%\noalign{\medskip}
\hline%\hline
%\noalign{\smallskip}
Group/Cluster & $<\!v_{220}\!>$ & \multicolumn{1}{c}{$\qquad\mu^0$} & Method \\
%\noalign{\smallskip}
\hline
%\noalign{\smallskip}
South Polar gr.$^*$ &  $112$ & $26.68\pm0.20$    & Cepheids (1)$^{\rm a}$ \\
M\,101 gr.$^*$      &  $405$ & $29.36\pm0.10$    & Cepheids (1)$^{\rm a}$,
                                                   TRGB$^{\rm f}$ \\
Leo$^*$             &  $652$ & $30.21\pm0.05$    & SNe\,Ia (2)$^{\rm b}$, 
                             Cepheids (3)$^{\rm a}$, TRGB$^{\rm f}$ \\
UMa$^{\dagger}$     & $1060$ & $\ge31.33\pm0.15$ & TF$^{\rm c}$, 
                                                   LC$^{\rm d}$ \\
Virgo$^{\ddagger}$  & $1179$ & $31.60\pm0.20$    & SNe\,Ia (3)$^{\rm b}$,
                                    Cepheids (4)$^{\rm a}$, TF$^{\rm
                                      c}$, LC$^{\rm d}$, GC$^{\rm e}$ \\
Fornax$^*$          & $1338$ & $31.60\pm0.10$    & SNe\,Ia (3)$^{\rm b}$, 
                                                   Cepheids (3)$^{\rm a}$\\ 
Coma                & $7188$ & $35.31\pm0.22$    & $3\mag71\pm0.08$ more
distant than Virgo$^{\rm g}$ \\
%\noalign{\smallskip}
\hline
%\noalign{\smallskip}
\end{tabular} %\footnotesize

$^*$ Group/cluster membership as defined by Kraan-Korteweg (1986)

$^{\dagger}$ Two separate agglomerations can be distinguished in the
UMa cluster; the nearer, presumably more nearly complete agglomeration
with 18 TF distances is considered here (Federspiel 1999)

$^{\ddagger}$ Cluster membership as defined by Binggeli et~al. (1993)

Sources: 
(a) For a list of individual Cepheid distances see Tammann
  et~al.\ (2001) (the number of Cepheid distances is shown in
  parentheses); 
(b) SNe\,Ia calibration as in Section~2.3 (the number of
  SNe\,Ia is shown in parentheses); 
(c) Federspiel (1999); 
(d) Luminosity classes (Sandage 2001); 
(e) Globular Clusters (Tammann \& Sandage 1999); 
(f) Kennicutt et~al.\ (1998); 
(g) Tammann \& Sandage (1999) 
\end{minipage}
\end{center}
\end{table}
% ******************************************************************

   The seven groups/clusters within $v_{220}<1200\kms$ in
Table~\ref{tab:2} define a Hubble diagram with $H_{0}=57.5\pm2.1$
(giving half weight to the nearby South Polar group) with a scatter of
only $\sigma_{(m-M)}=0.17$. If also the Coma cluster is included one
obtains $H_0=58.2 \pm 1.8$, $\sigma_{(m-M)}=0.16$.

% **************************************************************
  \subsection{Summary}\label{sec:2_5}
% **************************************************************
%
The distances discussed in sections 2.1 to 2.4 are plotted in a
distance calibrated Hubble diagram (Fig.~\ref{fig:3}). They are best
fit by a local value of $H_0(<1200\kms)=59.2\pm1.4$ and a large-scale
value out to $\sim30\,000\kms$ of $H_0({\rm cosmic})=57.4\pm2.3$. The
difference between the local and cosmic value of $H_0$ is
insignificant. This demonstrates that the expansion rate changes very
little with scale. 
% ******************************************************************
%  Figure 3
% ******************************************************************
\def\floatwidth{1.0\textwidth}
\begin{figure}%[tbh]
  \centerline{\includegraphics[width=\floatwidth]{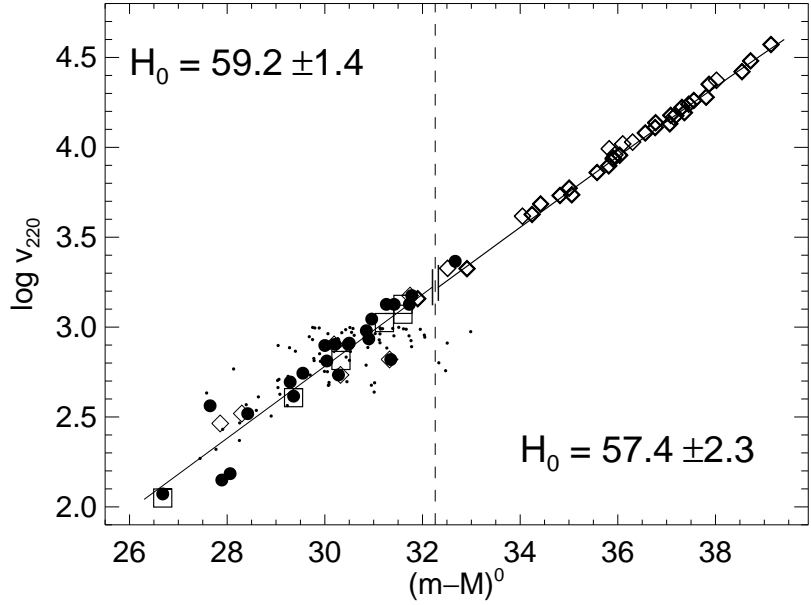}}
\caption{A synoptic distance-calibrated Hubble diagram extending over
  11 magnitudes or a factor of 100 in recession velocity. The
  left-hand value of $H_0$ based on Cepheids (large dots), TF
  distances (small dots), and groups/clusters (squares) is independent
  of the right-hand value of $H_0$ based on SNe\,Ia (diamonds), except
  that all distances rest ultimately on Cepheids. The full drawn line
  corresponds to $\Omega_{\rm Matter}=0.3$, $\Omega_{\Lambda}=0.7$.} 
\label{fig:3}
\end{figure}
% ******************************************************************

% **************************************************************
  \section{Desiderata}\label{sec:3}
% **************************************************************
%
One could gain the impression from Section~2 that the problem of
extragalactic distances was solved. Nothing is further from the
truth. Only about 200 galaxies with random individual distance errors
between $0\mag1$ and $0\mag5$ were involved. For the vast majority of
galaxies one has still no distances. Of course, a dynamical distance
can be obtained from the recession velocity and an adopted value of
$H_0$, but the accuracy is limited by peculiar motions. If one assumes
$630\kms$ as a typical large-scale peculiar motion, as judged by the
local motion with respect to the microwave background, the dynamical
distances of galaxies beyond $10\,000\kms$ are affected by only $\la6\%$
from peculiar motions. For galaxies at $3000\kms$ the dynamical
distance error is roughly tripled, but stays at this value for still
nearer galaxies because the peculiar motions tend to decrease over
shorter scales. The ``Hubble'' scatter of the Cepheid distances
suggests that $v_{\rm pec}/v_{220}\sim0.2$ (cf. Section 2.1.3). Of
course, the dynamical distances of very nearby galaxies
($v_{220}\la200\kms$ or $\la4\;$Mpc) are highly unreliable, and
they require ``honest'' distance determinations.

   This picture is over-optimistic for galaxies in high-density
regions (like the Virgo region) because the peculiar motions are
larger here.

   Some investigations depend just on a comparison of dynamical
distances and ``honest'' distance determinations, e.g.\ if the infall
into the Virgo cluster should be studied, or generally if the local
and not so local expansion field should be mapped, which would result,
among other things, in a delineation of the co-moving volume which
partakes in the local motion towards the microwave background.  

   Just for this purpose the data in Fig.~\ref{fig:3} are insufficient
because there are only four SNe\,Ia in the most relevant range
$1200<v_{220}<10\,000\kms$ and not a single SNe\,Ia in the range
$2500<v_{220}<4000\kms$! For the velocity mapping {\em relative\/} TF
and D$_{\rm n}-\sigma$ distances of clusters can close the
gap. Giovanelli et~al.\ (1997) and Dale et~al.\ (1999) have derived TF
distances of 71 clusters out to $v<25\,000\kms$. They define a tight
Hubble line with a scatter ($\sigma_{(m-M)}=0.11$) rivalling
SNe\,Ia. But they do not yield an independent value of $H_0$; this is
because the individual cluster distances are the mean of about ten
subjectively selected cluster members. They are far from constituting
a complete sample, which is the {\em conditio sine qua non\/} for any
application of equation~(\ref{eq:14}). Also the relative D$_{\rm
  n}-\sigma$ distances of ten clusters by Kelson et~al.\ (2000) cannot
independently be put on an absolute scale for the same reason
(cf. Tammann 2001). 

   In the preceding pages only random statistical errors were quoted,
although systematic errors are probably prevailing. Most serious are
of course systematic errors of the P-L relation of Cepheids because
the local expansion field on the left side of Fig.~\ref{fig:3} as well
as the large-scale expansion field on the right side depend on the
goodness of the calibrating Cepheid distances.

% **************************************************************
  \subsection{Systematic errors of the Cepheid distances}\label{sec:3_1}
% **************************************************************
%
% **************************************************************
  \subsubsection{The shape of the P-L relation}\label{sec:3_1_1}
% **************************************************************
%
So far all Cepheid distances are based on equations~(\ref{eq:2}) and
(\ref{eq:3}). If equations~(\ref{eq:10}) and (\ref{eq:11}) were used
instead they would be reduced by $0\mag13$ or 6\%
(equation~\ref{eq:12}) on average, assuming a median period of
$30^{\rm d}$ for the Cepheids involved.  
Yet the new P-L relations are only defined up to
$30^{\rm d}$, the OGLE images being saturated for LMC Cepheids with
longer periods. Identification and photometry of the rare Cepheids
with long periods are therefore still needed. 

% **************************************************************
  \subsubsection{The zero point of the P-L relations}\label{sec:3_1_2}
% **************************************************************
%
The distance modulus $(m-M)_{\rm LMC}=18.56$ (Table~\ref{tab:1}),
which defines the zero point of the P-L relations, has a small
statistical error, but a systematic error of about $\pm0\mag08$ cannot
be excluded. The whole extragalactic distance scale would benefit from
a firm LMC modulus.

% **************************************************************
  \subsubsection{The metallicity effect of Cepheids}\label{sec:3_1_3}
% **************************************************************
%
No corrections for the dependence of Cepheid luminosities on
metallicity have been applied, because even the sign of these
corrections is not well known (cf. Tammann et~al.\ 2001). Model
calculations (Chiosi et~al.\ 1993; Saio \& Gautschy 1998; Sandage
et~al.\ 1999; Alibert \& Baraffe 2000) as well as observations
(Kennicutt et~al.\ 1998; Tammann et~al.\ 2001) indicate that the
effect is smaller than $\Delta M = \pm 0.25 \Delta$\,[Fe/H], 
yet a dependence of this size would still
decrease/increase the known Cepheid distances by $0\mag08$ (4\%) on
average. Independent distance information of Cepheids of different
metallicity are therefore important.

% **************************************************************
  \subsubsection{The width of the instability strip}\label{sec:3_1_4}
% **************************************************************
%
The finite width of the instability strip in the color-magnitude
diagram causes an {\em intrinsic\/} scatter of the P-L relation which
decreases with increasing wavelength (see the scatter in
Fig.~\ref{fig:1}). 
A P-L relation gives therefore the absolute magnitude and distance
of a single Cepheid to only $\sim\!\pm0\mag3$. A P-L-color (P-L-C)
relation of the form 
\begin{equation}\label{eq:18a}
   M = a \log P + b(B-V) + c
\end{equation}
has repeatedly been proposed to position a Cepheid within the
instability strip. However, the pulsation models of Saio \& Gautschy
(1998) have shown that the constant-period lines in the instability
strip change slope in function of mass, and hence the coefficient $b$
becomes a function of period. It has also been proposed to use the
amplitude to locate a Cepheid within the strip (Sandage \& Tammann
1971). Lack of sufficient data has brought these attempts to a
preliminary halt.

% **************************************************************
  \subsection{Difficulties with RR Lyrae Stars}\label{sec:3_2}
% **************************************************************
%
As stated before the calibration of the terms $a$ and $b$ in
equation~(\ref{eq:13}) is difficult, if not to say
controversial. Particularly statistical parallaxes of Galactic RR
Lyrae stars, which strongly depend on the sample definition, have
confused the issue. But there remains a tendency of RR\,Lyr's ---
although sometimes exaggerated --- to yield a smaller LMC modulus than
Cepheids (cf. Table~\ref{tab:1}). There is still the possibility that
an additional parameter influences the RR\,Lyr luminosities.

% **************************************************************
  \subsection{Securing the TRGB Calibration}\label{sec:3_3}
% **************************************************************
%
The luminosity calibration of the $I$-band TRGB rests so far only on a
few adopted globular cluster distances, or otherwise has to rely on
the Cepheid distances of external galaxies. Consequently the
insensitivity of the TRGB against metallicity is not well tested
yet. A definitive calibration would make the TRGB a powerful tool
requiring relatively little observing time.

% **************************************************************
  \subsection{Improving the SNe\,Ia Calibration}\label{sec:3_4}
% **************************************************************
%
The large-scale value of $H_0$ depends entirely on the absolute
magnitudes of SNe\,Ia (equation~\ref{eq:17}), whose goodness in turn
depends on nine calibrating Cepheid distances. Any improvement of the
P-L relation of Cepheids reflects therefore directly on
$H_0$\,(cosmic). --- Additional SNe\,Ia at intermediate distances are
needed to tighten the correlation of their absolute magnitudes
$M_{\max}$ with their decline rate $\Delta m_{15}$ and intrinsic color
$(B-V)$. Allowance for this correlation only develops the full power
of SNe\,Ia as standard candles, i.e. it reduces the observed scatter
of their Hubble diagram from $0\mag17$ to $0\mag11$. But the
correction of $M_{\max}$\,(observed) for second parameters has also a
systematic effect on $H_0$. This is because the nine calibrating
SNe\,Ia lie necessarily in late-type spirals (to contain Cepheids),
and SNe\,Ia in spirals tend to be brighter and bluer and to have
smaller decline rates than average (cf. Parodi et~al.\ 2000).

% **************************************************************
\section{GAIA's Impact on the Extragalactic Distance Scale}\label{sec:4}
% **************************************************************
%
% **************************************************************
  \subsection{Distances of nearby Galaxies}\label{sec:4_1}
% **************************************************************
%
GAIA will measure the trigonometric parallaxes of countless stars in
LMC and SMC as well as in UMi and in at least seven other nearby dwarf
galaxies. This will not only provide fundamental distances of these
galaxies but also fix their orientation in space.

   GAIA will in addition provide rotational parallaxes for some nearby
galaxies like M\,31 and M\,33 and marginally even for galaxies outside
the Local Group like NGC\,55, NGC\,247, NGC\,253, NGC\,300, M\,81 and
NGC\,2403, by comparing their rotation curves observed from radial
velocities (probably best observed from the ground) and from proper
motions.  

   The direct distance measurements of nearby galaxies --- important
for themselves --- will offer a control of the distance indicators
discussed in the next Section.

% **************************************************************
  \subsection{The Calibration of Distance Indicators}\label{sec:4_2}
% **************************************************************
%
GAIA will determine trigonometric parallaxes of thousands of Galactic
Cepheids, RR~Lyrae stars, and of stars at the TRGB. It will also
provide highly reliable ZAMS positions in function of metallicity, and
the position of the cooling sequence of White Dwarfs (WD). 
Moreover, GAIA will measure the parallaxes of most of the 140 Galactic
Globular Clusters (GC) to better than 1\%. Their luminosity function
(GCLF) may in turn become valuable for an extragalactic distance scale
based on only old objects (cf.\ 2.2.1). Finally GAIA will decide to
what extent the P-L relation of Mira variables can be made useful for
distance determinations.  

\begin{figure}[t]
\begin{center}
{\footnotesize \sf
\def\floatwidth{0.25\textwidth}
\setlength{\unitlength}{1mm}
\def\xx{100}
\def\yy{100}
\begin{picture}(\xx,\yy)(0,0)
% **************************************************************
 \put(0,0){\line(\xx,0){\xx}}
 \put(0,0){\line(0,\yy){\yy}}
 \put(0,\yy){\line(\xx,0){\xx}}
 \put(\xx,0){\line(0,\yy){\yy}}
% **************************************************************

\def\px{20}
\def\py{70}
\put(\px,\py){\circle{40}}
\put(\px,\py){\makebox(0,0)[c]{\parbox{1cm}{\centering \scriptsize
      \,TRGB$^*$ \\ stars}}}
\def\py{50}
\put(\px,\py){\circle{40}}
\put(\px,\py){\makebox(0,0)[c]{\parbox{1cm}{\centering \scriptsize
      \,ZAMS$^*$ \\ stars}}}
\def\py{30}
\put(\px,\py){\circle{40}}
\put(\px,\py){\makebox(0,0)[c]{\parbox{1cm}{\centering \scriptsize
      White \\ Dwarfs }}}
\def\px{80}
\def\py{60}
\put(\px,\py){\circle{40}}
\put(\px,\py){\makebox(0,0)[c]{\parbox{1cm}{\centering \scriptsize
      RR\,Lyr$^*$}}}
\def\py{40}
\put(\px,\py){\circle{40}}
\put(\px,\py){\makebox(0,0)[c]{\parbox{1cm}{\centering \scriptsize
      Globular \\ Clusters}}}
\def\px{50}
\def\py{80}
\put(\px,\py){\circle{40}}
\put(\px,\py){\makebox(0,0)[c]{\parbox{1cm}{\centering \scriptsize
      $\!\!$Cepheids$^*$}}}
\def\py{20}
\put(\px,\py){\circle{40}}
\put(\px,\py){\makebox(0,0)[c]{\parbox{1cm}{\centering \scriptsize
      LMC \\ SMC \\ +others}}}
\put(50,50){\vector(0,0){23}}
\put(50,50){\vector(0,-1){23}}
\put(80,50){\vector(0,0){3}}
\put(80,50){\vector(0,-1){3}}
\put(28,28.5){\vector(4,1){45}}
\put(28,28.5){\vector(-4,-1){1}}
\put(28,51.5){\vector(4,-1){45}}
\put(28,51.5){\vector(-4,1){1}}
\put(24.5,64.5){\vector(2,-3){25}}
\put(24.5,64.5){\vector(-2,3){0}}
\put(27,71.7){\vector(3,-2){46}}
\put(27,71.7){\vector(-3,2){0}}
\put(73,59.5){\vector(-2,-3){21.7}}
\put(73,59.5){\vector(2,3){0}}
%
% **************************************************************
\end{picture}
}
  \end{center}
  \caption{Objects for which GAIA will determine trigonometric
    parallaxes and which in turn are important for the distance scale
    are shown inside circles with arrows indicating their
    interdependencies. Stars denote distance indicators whose
    metallicity dependencies will be also determined by GAIA.}
  \label{fig:4}
\end{figure}
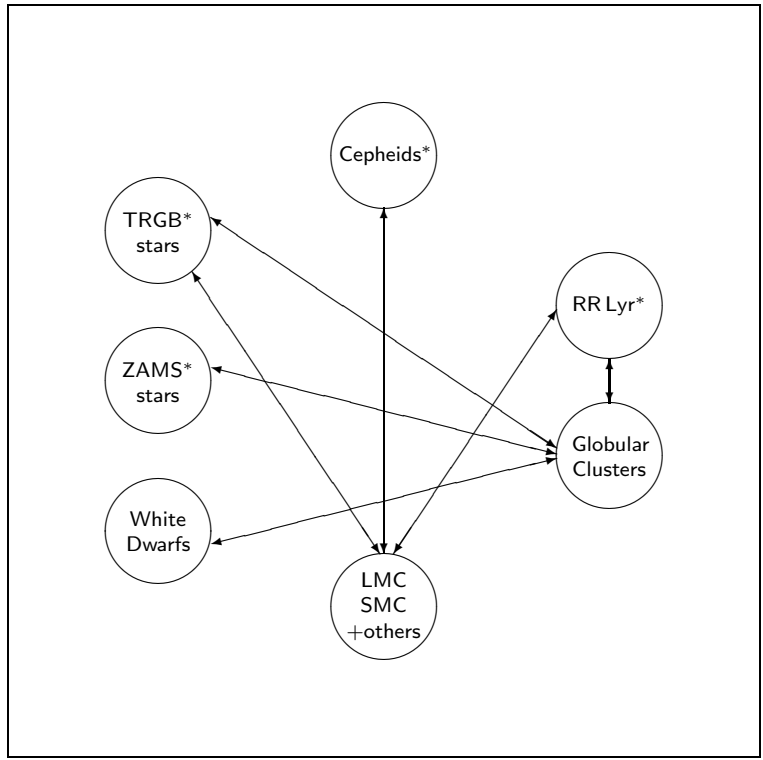

   The above distance indicators are interlocked in a multiple way
(Fig.~\ref{fig:4}). RR~Lyrae and TRGB stars as well as ZAMS and WD
fitting will yield distances of GCs whose trigonometric parallaxes
will be known. 
Distances to LMC, SMC, and other nearby galaxies with known
trigonometric and some rotational parallaxes will be measured again by
Cepheids and RR~Lyrae and TRGB stars. This network of independent
distance determinations will clarify the metallicity dependence of
each method. Any discrepancies which may occur will lead to new
astrophysical insights. 
% **************************************************************
%           Figure 4.
% **************************************************************
%\clearpage
% **************************************************************
% **************************************************************

% **************************************************************
  \subsection{Intermediate Galaxy Distances from SNe\,Ia}\label{sec:4_3}
% **************************************************************
%
GAIA offers a unique opportunity to discover ``nearby'' SNe\,Ia
($v<10\,000\kms$). Each SN\,Ia with good photometry yields the chance
of the century to determine the distance of its host galaxy to within
$0\mag11$ (5\%). No other distance indicator can give the same
accuracy, --- except galaxies with $v>10\,000\kms$ for which the best
distance is obtained from its recession velocity and by assuming a
(correct) value of $H_0$ (cf.\ Section~3). 

   Therefore SNe\,Ia with $v\le10\,000\kms$ are of particular
interest. Out to this limit their number can be estimated from their
frequency per unit luminosity and from the luminosity density summed
over all galaxies (both scaled to [for instance] $H_0=60$). 
The former is $\sim\!0.22$ SNe\,Ia per $10^{10}$ solar luminosities
(${L}_{\odot}$) and per 100 years, independent of the galaxy type
(Tammann et~al. 1994); 
the total galaxian luminosity density is $1.3 \cdot
10^{8}L_{\odot}\;$Mpc$^{-3}$ (Tammann 1982). The volume $V
(<10\,000\kms)$ being $19.4\cdot 10^6\;$Mpc$^3$ yields then 555
SNe\,Ia per year with an error of probably less than a factor 2.
All of them have $m_{\max}^{B,V}\la16.6\pm\sigma_{M}$, where
$\sigma_{M}=0\mag2$ for SNe\,Ia {\em before\/} correction for decline
rate and color. Half of these SNe\,Ia will
be caught by GAIA, which covers every point in the sky once per month,
near maximum, i.e.\ earlier than 10 days after maximum. At this time a
SN\,Ia has faded to $m^{B,V}\la17.0$ assuming the standard light curve
of Leibundgut (1991), and if it is subsequently followed for 30
days, even the faintest SN\,Ia of the sample will have
$m^{B,V}\la19.0$. This follow-up is sufficient to extra\-polate
$m_{\max}^{B,V}$ and to determine the decline rate $\Delta m_{15}$ and
$(B-V)_{\max}$ with sufficient accuracy even for SNe\,Ia whose maxima
have been missed by 10 days. The resulting $m_{\max}^{B,V}$ corrected
for $\Delta m_{15}$ and $(B-V)_{\max}$ will yield, if combined with
$M_{\max}^{B,V}$ in equation~(\ref{eq:17}), unrivalled distances for
1100 galaxies during the 4-year operation time of GAIA.

   $\!\!\!\!\!$ This number is overoptimistic because of Galactic
absorption. Roughly one third of the SNe\,Ia will be at low Galactic
latitudes, i.e.\ $|b|<20^{\circ}$ and be dimmed by
$A_{V}\ga0\mag4$. They are either missed, or their photometry is
impaired by large absorption corrections. SNe\,Ia at higher $|b|$
will suffer an absorption of $A_{V}\sim0\mag25$ on average (cf.\
Schlegel et~al.\ 1998). A corresponding margin of the detection and
follow-up limits should therefore be allowed for.
Moreover, many day time SNe\,Ia will not be suitable for follow-up
from the ground.

   Nevertheless the benefit of this programme is fourfold: 

\noindent
(1) The precision distances of these galaxies will map the not so
    local velocity field and explain the origin of the CMB dipole in
    great detail;

\noindent
(2) The residuals of the new SNe\,Ia from the mean Hubble line will
    fix the dependencies of $m_{\max}^{B,V}$ on $\Delta m_{15}$ and
    $(B-V)_{\max}$ with any desiderable accuracy. These dependencies
    rest now only on 35 SNe\,Ia and their errors can still cause
    systematic errors of all SNe\,Ia distances of 3-4\% (cf. Parodi
    et~al.\ 2000).

\noindent
(3) The new SNe\,Ia will richly populate the gap of the Hubble diagram
    (right side of Fig.~\ref{fig:3}) and fix the position of the
    Hubble line to within a vanishingly small error, reducing still
    further the random error of the constant term $c$ in
    equation~(\ref{eq:16}) and correspondingly also of $H_0$. Moreover,
    the position of the Hubble line at $v\la10\,000\kms$ is essential
    when the cosmological constant $\Lambda$ shall be derived from the
    Hubble diagram of very distant SNe\,Ia.

\noindent
(4) The new SNe\,Ia, for which one will have reliable survey times,
    are ideally suited for the determination of their frequency with
    unprecedented accuracy.

GAIA cannot provide the follow-up photometry of the newly discovered
SNe\,Ia. The above SNe\,Ia program is therefore only meaningful if it
will be possible to organize a dedicated (earth-bound) program for the
photometry in $B$, $V$ (and $I$) of the new SNe\,Ia down to $\sim\!
19\mag0$.

If one wants to carry the GAIA search for SNe\,Ia to still fainter
limits their total number per year increases as 
\begin{equation}\label{eq:19}
   \log N_{\rm SNeIa} = 2.44 + 0.6 (m_{\rm limit} - 17.0)
\end{equation}
(cf. H{\o}g et~al.\ 1999).

   The SN sample provided by GAIA will be little ``contaminated'' by
SNe\,II/Ib because they are fainter by $2\mag5$ on average than
SNe\,Ia. Therefore all magnitude-limited SN searches are strongly
biased against SNe\,II/Ib in spite of their considerably higher
frequency per unit volume. Their value for cosmology is much smaller
because their luminosity function is broad ($\sigma_{M}=1\mag2$;
Tammann \& Schr{\"o}der 1990) and even skewed towards fainter SNe\,II
(Woltjer 1998). They can hence not be used as standard candles. The
determination of their {\em individual\/} distances from the expanding
photosphere method (EPM) poses formidable problems due to the Compton
scattering in {\em  moving\/} atmospheres. Available results may
therefore still be affected by systematic effects (cf. Hamuy et~al.\
2001 and references therein). 

   GAIA's high resolution power will reveal a large number of
gravitationally lensed double or multiple quasars. In principle
they offer themselves for an independent distance determination,
but the solution is typically degenerate as to the distance {\em
  and\/} the lensing mass. 
Thus a single lensed quasar leaves a wide margin for $H_{0}$ (Saha \&
Williams 2001). A {\em combined\/} solution for six quasars gives
$50<H_{0}<60$ (Saha 2002). This value can be much tightened when
additional cases will be discovered.

% **************************************************************
  \section{Conclusions}\label{sec:5}
% **************************************************************
%
Reliable distance indicators (Cepheids, TF distances of complete
samples, and nearby cluster distances) require a {\em local\/} value
of $H_0=59.2\pm1.4$. Cepheid-calibrated SNe\,Ia give a {\em
  large-scale\/} value of $H_0=57.4\pm2.3$. The two values are
statistically indistinguishable and one may assume $H_0=58\pm2$
everywhere. 

   The quoted error is only the statistical error. But the
solution for $H_0$ is dominated by {\em systematic\/} errors which
could amount to as much as 10\%. The largest systematic errors are
introduced by Cepheids, which form the basis of the local as well as
of the large-scale expansion rate. Equation~(\ref{eq:12}) suggests, in
case of a non-linear form of the P-L relation of Cepheids, that $H_0$
is increased by 6\% for Cepheids with a median period of 30 days,
i.e.\ a typical period for extra\,-\,Local Group galaxies. A metallicity
dependence of the Cepheid P-L relation could affect $H_0$ by $\sim\!5\%$; if
the metallicity correction of Kennicutt et~al.\ (1998) is taken at face
value the effect would go in the direction of decreasing $H_0$. The
zero point of the P-L relation seems to be well determined
(Table~\ref{tab:1}), but a systematic error of 3-4\% cannot be
excluded. 

   The three systematic error sources of Cepheid distances, which
propagate with almost full weight into the entire extragalactic
distance scale, will essentially be eliminated by GAIA.

The next important source of a systematic error --- at least for the
large-scale value of $H_0$ --- comes from the homogenization of SNe\,Ia
as to decline rate $\Delta m_{15}$ and color $(B-V)$. In fact errors
of the slope of the $M-\Delta m_{15}$ and $M-(B-V)$ relations could
introduce systematic errors of the SNe\,Ia distances of $\sim 3\%$
(Parodi et~al.\ 2000). This problem can entirely be solved if good
photometry will be obtained for the many hundreds of SNe\,Ia to be
discovered within $10\,000\kms$ by GAIA. This at the same time 
will yield irreplaceable distances to an equal number of field
galaxies (to within random errors of $\pm5\%$), which will outline
any concerted deviations from pure Hubble flow out to $10\,000\kms$.   

   The one remaining systematic error of $\la 4\%$ comes from the
difficult photometry of extra-Local Group Cepheids with the wide-field
camera (WFPC-2) of HST. GAIA cannot offer a handle on this problem,
and it has to await future photometry from space.

   GAIA will not only much improve the value of $H_0$, but also the
age of the oldest objects. Definitive distances to globular clusters
will reduce the error of their ages, and the improvement of the
distances of Local Group galaxies {\em and\/} the measurement of their
proper motions are very important for the determination
of the dynamical age of the Local Group (Lynden-Bell 1999). Thus $H_0$
and a minimum age of the Universe can be used as strong priors for the
CMB fluctuation spectrum (cf. Netterfield et~al.\ 2002; Pryke 2001) 
which will narrow down the possible range of other cosmological
parameters like the baryon density $\Omega_{\rm b}$, the matter
density $\Omega_{\rm Matter}$, and the cosmological constant $\Lambda$. 

\begin{acknowledgements}
The authors thank the Swiss National Science Foundation for financial
support. 
\end{acknowledgements}

\noindent
{\em Note added in proof (Sep.~5, 2002).} It has become clear now that
the P-L relation of Galactic fundamental-mode Cepheids is distinctly
different from that in LMC.
The Galactic relation is quite steep, the short-period Cepheids being
fainter than in LMC and the long-period ones being brighter (Tammann,
Sandage, \& Reindl, in preparation). The non-uniqueness of the P-L
relations --- which is probably due, at least in part, to metallicity
differences --- greatly increases the need for fundamental distances
of nearby galaxies from GAIA.

% ******************************************************************
% Bibliography
% ******************************************************************

\end{article}
\end{document}